\documentstyle[amssymb,floats,aps,prb,epsf]{revtex}
\epsfclipon
\begin{document}
\twocolumn[\hsize\textwidth\columnwidth\hsize\csname
@twocolumnfalse\endcsname

\draft
\draft
\title{Unusual heat transport in underdoped cuprates}

\author{Tianxing Ma and Shiping Feng}
\address{Department of Physics and Key Laboratory of Beam
Technology and Material Modification, Beijing Normal University,
Beijing 100875, China}

\maketitle
\begin{abstract}

Within the $t$-$J$ model, the heat transport of the underdoped
cuprates is studied based on the fermion-spin theory. It is shown
that at low temperatures the energy dependence of the thermal
conductivity spectrum consists of two bands. The higher-energy
band shows a weak peak, while the low-energy peak is located at a
finite energy. This high-energy broad band is severely suppressed
with increasing temperatures, and vanishes at higher temperature.
It is also shown that the temperature dependence of the thermal
conductivity increases monotonously with increasing temperatures,
in agreement with experiments.
\end{abstract}
\pacs{74.25.Fy,74.62.Dh,72.15.Eb}

]
\bigskip

\narrowtext

It has become clear in the past fifteen years that the cuprate
superconductors are indeed fundamentally different from other
conventional metals in that they are doped Mott insulators
\cite{kbse1,bcmps}. The undoped cuprates are Mott insulators with
the antiferromagnetic (AF) long-range-order (AFLRO). A small
amount of carrier dopings to this Mott insulating state drives the
metal-insulator transition and directly results in the
superconducting transition at low temperatures for low carrier
dopings \cite{kbse1,bcmps}. The optical conductivity in the normal
state above superconducting transition temperature shows a
non-Drude behavior at low energies, and is carried by $x$ holes,
with $x$ is the hole doping concentration, while the resistivity
exhibits a linear temperature behavior over a wide range of
temperatures in the underdoped regime \cite{cooper1}. This unusual
charge transport does not fit in the conventional Fermi-liquid
theory \cite{pwa1}. Furthermore, a clear departure from the
universal Wiedemann-Franz law for the typical Fermi-liquid
behavior is observed in doped cuprates \cite{hill}. It has been
argued that these anomalous features may be interpreted within the
framework of the charge-spin separation \cite{pwa1,pwa2,houghton},
where the electron is separated into a neutral spinon and charged
holon, therefore the basic excitations of doped cuprates are not
fermionic quasiparticles as in other conventional metals with
charge, spin and heat all carried by one and the same particles.

The heat transport, as manifested by the thermal conductivity, is
one of the basic transport properties that provides a wealth of
useful informations on the carriers and phonons as well as their
scattering processes \cite{uher,nakamura,sun}. In the conventional
metals, the thermal conductivity contains both contributions from
carriers and phonons \cite{uher}. The phonon contribution to the
thermal conductivity is always present in the conventional metals,
while the magnitude of the carrier contribution depends on the
type of material because it is directly proportional to the free
carrier density. In particular, the free carrier density in
conventional superconducting materials is very high, then the
thermal conductivity from the free carriers is usually the
dominant contribution. In the cuprate materials, the physical
properties in the undoped case are now quite well understood
\cite{kbse1}: here the system of interacting localized Cu$^{2+}$
spins is well described as the conventional insulating
antiferromagnet, and at low temperatures, it has the purely phonon
thermal conductivity \cite{nakamura}. However, it has been shown
from experiments \cite{nakamura,baberski} that the phonon
contribution to the thermal conductivity is strongly suppressed in
the underdoped regime, and the conventional models of phonon heat
transport based on phonon-defect scattering or conventional
phonon-electron scattering fail to explain the experimental data
\cite{baberski}. Since the electron has been separated as the
holon and spinon, and therefore it has been argued that the
contributions from spinons and holons may dominate the heat
transport of the underdoped cuprates \cite{baberski,houghton}.
Recently, we \cite{feng1} have developed a fermion-spin theory
based on the charge-spin separation to study the physical
properties of doped cuprates, where the electron operator is
decoupled as the gauge invariant dressed holon and spinon. Within
this theory, we have discussed charge transport and spin response
of the underdoped cuprates and superconducting mechanism. It has
been shown that the charge transport is mainly governed by the
scattering from the dressed holons due to the dressed spinon
fluctuation, while the scattering from the dressed spinons due to
the dressed holon fluctuation dominates the spin response
\cite{feng1,feng2}. These dressed holons interact occurring
directly through the kinetic energy by exchanging the dressed
spinon excitations, leading to a net attractive force between the
dressed holons, then the electron Cooper pairs originating from
the dressed holon pairing state are due to the charge-spin
recombination, and their condensation reveals the superconducting
ground-state, where the electron superconducting transition
temperature is determined by the dressed holon pair transition
temperature, and is proportional to the hole doping concentration
in the underdoped regime \cite{feng3}, in agreement with the
experiments. In this paper, we apply this successful approach to
discuss the heat transport of the underdoped cuprates. Within the
$t$-$J$ model, we show that although both dressed holons and
spinons are responsible for the heat transport of the underdoped
cuprates, the contribution from the dressed spinons dominates the
thermal conductivity.

In the doped cuprates, the single common feature is the presence
of the two-dimensional (2D) CuO$_{2}$ plane \cite{kbse1}, then it
is believed that the unusual physical properties are closely
related to the doped CuO$_{2}$ planes. It has been argued
\cite{pwa2} that the essential physics of the doped CuO$_{2}$
planes is contained in the 2D $t$-$J$ model,
\begin{eqnarray}
H&=&-t\sum_{i\hat{\eta}\sigma}C^{\dagger}_{i\sigma}
C_{i+\hat{\eta}\sigma}+\mu \sum_{i\sigma}
C^{\dagger}_{i\sigma}C_{i\sigma} \nonumber \\
&+&J\sum_{i\hat{\eta}}{\bf S}_{i}\cdot {\bf S}_{i+\hat{\eta}},
\end{eqnarray}
with $\hat{\eta}=\pm\hat{x},\pm\hat{y}$, $C^{\dagger}_{i\sigma}$
($C_{i\sigma}$) is the electron creation (annihilation) operator,
${\bf S}_{i}=C^{\dagger}_{i}{\vec\sigma}C_{i}/2$ is spin operator
with ${\vec\sigma}=(\sigma_{x},\sigma_{y},\sigma_{z})$ as Pauli
matrices, and $\mu$ is the chemical potential. The $t$-$J$ model
(1) is supplemented by the single occupancy local constraint
$\sum_{\sigma}C_{i\sigma}^{\dagger}C_{i\sigma}\leq 1$. This local
constraint reflects the strong electron correlation in the doped
Mott insulator \cite{pwa2}, and can be treated properly in
analytical form within the fermion-spin theory \cite{feng1} based
on the charge-spin separation,
\begin{eqnarray}
C_{i\uparrow}=h^{\dagger}_{i\uparrow}S^{-}_{i},~~~~
C_{i\downarrow}=h^{\dagger}_{i\downarrow}S^{+}_{i},
\end{eqnarray}
where the spinful fermion operator
$h_{i\sigma}=e^{-i\Phi_{i\sigma}}h_{i}$ describes the charge
degree of freedom together with some effects of the spinon
configuration rearrangements due to the presence of the hole
itself (dressed holon), while the spin operator $S_{i}$ describes
the spin degree of freedom (dressed spinon), then the electron
on-site local constraint for the single occupancy, $\sum_{\sigma}
C^{\dagger}_{i\sigma} C_{i\sigma}=S^{+}_{i}h_{i\uparrow}
h^{\dagger}_{i\uparrow}S^{-}_{i} +S^{-}_{i}h_{i\downarrow}
h^{\dagger}_{i\downarrow}S^{+}_{i}=h_{i} h^{\dagger}_{i}(S^{+}_{i}
S^{-}_{i}+S^{-}_{i}S^{+}_{i})=1- h^{\dagger}_{i}h_{i}\leq 1$, is
satisfied in analytical calculations, and the double spinful
fermion occupancy, $h^{\dagger}_{i\sigma}h^{\dagger}_{i-\sigma}
=e^{i\Phi_{i\sigma}} h^{\dagger}_{i}h^{\dagger}_{i}
e^{i\Phi_{i-\sigma}}=0$, $h_{i\sigma} h_{i-\sigma}=
e^{-i\Phi_{i\sigma}}h_{i}h_{i}e^{-i\Phi_{i-\sigma}}=0$, are ruled
out automatically. It has been shown that these dressed holon and
spinon are gauge invariant, and in this sense, they are real and
can be interpreted as the physical excitations
\cite{feng1,laughlin}. This dressed holon $h_{i\sigma}$ is a
spinless fermion $h_{i}$ incorporated a spinon cloud
$e^{-i\Phi_{i\sigma}}$ (magnetic flux), then is a magnetic
dressing. In other words, the gauge invariant dressed holon
carries some spinon messages, i.e., it shares its nontrivial
spinon environment due to the presence of the holon itself
\cite{martins}. Although in common sense $h_{i\sigma}$ is not a
real spinful fermion, it behaves like a spinful fermion. In this
fermion-spin representation, the low-energy behavior of the
$t$-$J$ model (1) can be expressed as \cite{feng1},
\begin{mathletters}
\begin{eqnarray}
H&=&\sum_{i}H_{i}, \\
H_{i}&=&-t\sum_{\hat{\eta}}(h_{i\uparrow}S^{+}_{i}
h^{\dagger}_{i+\hat{\eta}\uparrow}S^{-}_{i+\hat{\eta}}+
h_{i\downarrow}S^{-}_{i}h^{\dagger}_{i+\hat{\eta}\downarrow}
S^{+}_{i+\hat{\eta}}) \nonumber \\
&-&\mu\sum_{\sigma}h^{\dagger}_{i\sigma}h_{i\sigma}+J_{{\rm eff}}
\sum_{\hat{\eta}}{\bf S}_{i}\cdot {\bf S}_{i+\hat{\eta}},
\end{eqnarray}
\end{mathletters}
with $J_{{\rm eff}}=(1-x)^{2}J$, and $x=\langle
h^{\dagger}_{i\sigma}h_{i\sigma}\rangle=\langle h^{\dagger}_{i}
h_{i}\rangle$ is the hole doping concentration. As a consequence,
the kinetic energy ($t$) term in the $t$-$J$ model has been
expressed as the dressed holon-spinon interaction, which dominates
the essential physics of doped cuprates, while the magnetic energy
($J$) term is only to form an adequate spinon configuration.

Within the $t$-$J$ model (3), the thermal conductivity of doped
cuprates can be expressed as \cite{mahan},
\begin{eqnarray}
\kappa(\omega,T)=-{1\over T}{{\rm Im}\Pi_{Q} (\omega,T)\over
\omega},
\end{eqnarray}
with $\Pi_{Q}(\omega,T)$ is the heat current-current correlation
function, and is defined as,
\begin{eqnarray}
\Pi_{Q}(\tau-\tau')=-\langle T_{\tau}j_{Q}(\tau)j_{Q}(\tau')
\rangle,
\end{eqnarray}
where $\tau$ and $\tau'$ are the imaginary times, $T_{\tau}$ is
the $\tau$ order operator, while the heat current density is
obtained within the Hamiltonian (3b) by using Heisenberg's
equation of motion as \cite{mahan},
\begin{mathletters}
\begin{eqnarray}
j_{Q}&=&i\sum_{i,j}{\bf R}_{i}[H_{i},H_{j}]=j^{(h)}_{Q}+
j^{(s)}_{Q}, \\
j^{(h)}_{Q}&=&i(\chi t)^{2}\sum_{i\hat{\eta}\hat{\eta}'\sigma}
\hat{\eta}h_{i+\hat{\eta}'\sigma}^{\dagger}h_{i+\hat{\eta}\sigma}
+i\mu\chi t\sum_{i\hat{\eta}\sigma}\hat{\eta}
h_{i+\hat{\eta}\sigma}^{\dagger}h_{i\sigma},\\
j^{(s)}_{Q}&=&i{1\over 2}(\epsilon J_{\rm{eff}})^{2}
\sum_{i\hat{\eta}\hat{\eta}'}(\hat{\eta}-\hat{\eta}')[S^{+}_{i}
S^{-}_{i-\hat{\eta}+\hat{\eta}'}S^{z}_{i-\hat{\eta}}\nonumber \\
&+& S^{-}_{i-\hat{\eta}+\hat{\eta}'}S^{+}_{i}S^{z}_{i-\hat{\eta}}]
+i\epsilon J^{2}_{\rm{eff}}\sum_{i\hat{\eta}\hat{\eta}'}
(\hat{\eta}-\hat{\eta}')[S^{+}_{i}S^{-}_{i+\hat{\eta}}
S^{z}_{i+\hat{\eta}'} \nonumber \\
&-&S^{+}_{i}S^{-}_{i-\hat{\eta}}S^{z}_{i-\hat{\eta}+\hat{\eta}'}],
\end{eqnarray}
\end{mathletters}
where ${\bf R}_{i}$ is lattice site, $\epsilon=1+2t\phi/J_{\rm
eff}$, $\chi=\langle S_{i}^{+}S_{i+\hat{\eta}}^{-}\rangle$ is the
spinon correlation function, and $\phi=\langle
h^{\dagger}_{i\sigma}h_{i+\hat{\eta}\sigma}\rangle$ is the dressed
holon's particle-hole parameter. Although the heat current density
$j_{Q}$ has been separated into two parts $j^{(h)}_{Q}$ and
$j^{(s)}_{Q}$, with $j^{(h)}_{Q}$ is the holon heat current
density, and $j^{(s)}_{Q}$ is the spinon heat current density, the
strong correlation between dressed holons and spinons still is
considered self-consistently through the dressed spinon's order
parameters entering in the dressed holon's propagator, and the
dressed holon's order parameters entering in the dressed spinon's
propagator. In this case, the heat current-current correlation
function (5) can be calculated in terms of the full dressed holon
and spinon Green's functions $g_{\sigma}(k,\omega)$ and
$D(k,\omega)$. Following the discussions of the charge transport
\cite{feng1,feng2,bang}, we obtain the thermal conductivity of
doped cuprates as,
\begin{mathletters}
\begin{eqnarray}
\kappa(\omega,T)&=&\kappa_{h}(\omega,T)+\kappa_{s}(\omega,T), \\
\kappa_{h}(\omega,T)&=&-{1\over 2N}\sum_{k\sigma}\Lambda_{h}^{2}
\gamma^{2}_{sk}\int^{\infty}_{-\infty}{d\omega'\over 2\pi}
A_{h\sigma}(k,\omega'+\omega) \nonumber \\
&\times& A_{h\sigma}(k,\omega'){n_{F}(\omega'+\omega)-n_{F}
(\omega') \over T\omega}, \\
\kappa_{s}(\omega,T)&=&-{1\over 2N}\sum_{k}\Lambda_{s}^{2}
\gamma^{2}_{sk}\int^{\infty}_{-\infty}{d\omega'\over 2\pi}
A_{s}(k,\omega'+\omega) \nonumber \\
&\times& A_{s}(k,\omega'){n_{B}(\omega'+\omega)-n_{B}(\omega')
\over T\omega},
\end{eqnarray}
\end{mathletters}
where $\kappa_{h}(\omega,T)$ and $\kappa_{s}(\omega,T)$ are the
corresponding contributions from dressed holons and spinons,
respectively, $\gamma^{2}_{sk}=({\rm sin}^{2}k_{x}+{\rm sin}^{2}
k_{y})/4$, $\Lambda_{h}=Z\chi t(\mu-Z\chi t\gamma_{k})$,
$\Lambda_{s}= (ZJ_{{\rm eff}})^{2}\epsilon(2\epsilon\chi +2C-4\chi
\gamma_{k})$, $\gamma_{{\bf k}}=(1/Z)\sum_{\hat{\eta}}e^{i{\bf k}
\cdot \hat{\eta}}$, $Z$ is the number of the nearest neighbor
sites, the spinon correlation function $C=(1/Z^{2})
\sum_{\hat{\eta},\hat{\eta'}}\langle S_{i+\hat{\eta}}^{+}
S_{i+\hat{\eta'}}^{-}\rangle$, the dressed holon and spinon
spectral functions are obtained as $A_{h\sigma}(k,\omega)=-2{\rm
Im}g_{\sigma}(k,\omega)$ and $A_{s}(k,\omega)=-2{\rm Im}
D(k,\omega)$, respectively, while the full dressed holon and
spinon Green's functions have been discussed in detail in Ref.
\cite{feng1}, and can be expressed as $g^{-1}_{\sigma}
(k,\omega)=g^{(0)-1}_{\sigma}(k,\omega)-\Sigma_{h}(k,\omega)$ and
$D^{-1}(k,\omega)=D^{(0)-1}(k,\omega)-\Sigma_{s}(k,\omega)$, with
the mean-field (MF) dressed holon and spinon Green's functions,
$g^{(0)-1}_{\sigma}(k,\omega)=\omega-\xi_{k}$ and
$D^{(0)-1}(k,\omega)=(\omega^{2}-\omega_{k}^{2})/B_{k}$, and the
second-order dressed holon and spinon self-energies are obtained
by the loop expansion to the second-order \cite{feng1,feng2} as,
\begin{mathletters}
\begin{eqnarray}
&~&\Sigma_{h}({\bf k},\omega)={1\over 2}\left ({Zt\over
N}\right)^2 \sum_{pp'}(\gamma^{2}_{{\bf p'+p+k}}+\gamma^{2}_{{\bf
p'-k}})
{B_{p'}B_{p+p'}\over 4\omega_{p'}\omega_{p+p'}}\nonumber \\
&\times& \left ({F^{(h)}_{1}(k,p,p')\over \omega+\omega_{p+p'}-
\omega_{p'}-\xi_{p+k}}+{F^{(h)}_{2}(k,p,p')\over\omega+\omega_{p'}
-\omega_{p+p'}-\xi_{p+k}}\right. \nonumber \\
&+&\left.
{F^{(h)}_{3}(k,p,p')\over\omega+\omega_{p'}+\omega_{p+p'}
-\xi_{p+k}}-{F^{(h)}_{4}(k,p,p')\over\omega -\omega_{p+p'}-
\omega_{p'}-\xi_{p+k}}\right),\nonumber \\
\\
&~&\Sigma_{s}({\bf k},\omega)=\left ({Zt\over N}\right )^{2}
\sum_{pp'}(\gamma^{2}_{{\bf p'+p+k}}+\gamma^{2}_{{\bf p'-k}})
{B_{k+p}\over 2\omega_{k+p}} \nonumber \\
&\times& \left
({F^{(s)}_{1}(k,p,p')\over\omega+\xi_{p+p'}-\xi_{p'}
-\omega_{k+p}}-{F^{(s)}_{2}(k,p,p')\over
\omega+\xi_{p+p'}-\xi_{p'}
+\omega_{k+p}}\right ), \nonumber \\
\end{eqnarray}
\end{mathletters}
respectively, where $B_{k}=\lambda[2\chi^{z}(\epsilon \gamma_{{\bf
k}}-1)+\chi(\gamma_{{\bf k}}-\epsilon)]$, $\lambda=2ZJ_{eff}$, the
spinon correlation function $\chi^{z}=\langle
S_{i}^{z}S_{i+\hat{\eta}}^{z}\rangle$,
$F^{(s)}_{1}(k,p,p')=n_{F}(\xi_{p+p'})[1-n_{F}(\xi_{p'})]-n_{B}
(\omega_{k+p})[n_{F}(\xi_{p'})-n_{F}(\xi_{p+p'})]$, $F^{(s)}_{2}
(k,p,p')=n_{F}(\xi_{p+p'})[1-n_{F}(\xi_{p'})]+[1+n_{B}
(\omega_{k+p})][n_{F}(\xi_{p'})-n_{F}(\xi_{p+p'})]$, $F^{(h)}_{1}
(k,p,p')=n_{F}(\xi_{p+k})[n_{B}(\omega_{p'})-n_{B}(\omega_{p+p'})]
+n_{B}(\omega_{p+p'})[1+n_{B}(\omega_{p'})]$, $F^{(h)}_{2}(k,p,p')
=n_{F}(\xi_{p+k})[n_{B}(\omega_{p'+p})-n_{B}(\omega_{p'})]+n_{B}
(\omega_{p'})[1+n_{B}(\omega_{p'+p})]$, $F^{(h)}_{3}(k,p,p')=n_{F}
(\xi_{p+k})[1+n_{B}(\omega_{p+p'})+n_{B}(\omega_{p'})]+n_{B}
(\omega_{p'})n_{B}(\omega_{p+p'})$, $F^{(h)}_{4}(k,p,p')=n_{F}
(\xi_{p+k)}[1+n_{B}(\omega_{p+p'})+n_{B}(\omega_{p'})]-[1+n_{B}
(\omega_{p'})][1+n_{B}(\omega_{p+p'})]$, $n_{B}(\omega_{p})$ and
$n_{F}(\xi_{p})$ are the boson and fermion distribution functions,
respectively, and the MF dressed holon and spinon spectra are
given by $\xi_{k}=Zt\chi\gamma_{{\bf k}}-\mu$, and
$\omega^{2}_{k}=A_{1}(\gamma_{k})^{2}+A_{2}\gamma_{k}+A_{3}$,
respectively, with
$A_{1}=\alpha\epsilon\lambda^{2}(\epsilon\chi^{z}+\chi/2)$,
$A_{2}=-\epsilon\lambda^{2}[\alpha(\chi^{z}+\epsilon\chi/2)+
(\alpha C^{z}+(1-\alpha)/(4Z)-\alpha\epsilon\chi/(2Z))+(\alpha
C+(1-\alpha)/(2Z)-\alpha\chi^{z}/2)/2]$, $A_{3}=\lambda^{2}
[\alpha C^{z}+(1-\alpha)/(4Z)-\alpha\epsilon\chi/(2Z)+\epsilon^{2}
(\alpha C+(1-\alpha)/(2Z)-\alpha\chi^{z}/2)/2]$, and the spinon
correlation function
$C^{z}=(1/Z^{2})\sum_{\hat{\eta},\hat{\eta'}}\langle
S_{i+\hat{\eta}}^{z}S_{i+\hat{\eta'}}^{z}\rangle$. In order not to
violate the sum rule of the correlation function $\langle
S^{+}_{i}S^{-}_{i}\rangle=1/2$ in the case without AFLRO, the
important decoupling parameter $\alpha$ has been introduced in the
MF calculation \cite{kondo,feng4}, which can be regarded as the
vertex correction. All the above MF order parameters, decoupling
parameter $\alpha$, and chemical potential $\mu$ are determined by
the self-consistent calculation \cite{feng4}.

The thermal conductivity is one of the direct probes to observe
the low energy quasi-particles through its frequency and
temperature dependences \cite{uher,nakamura,sun}. In Fig. 1, we
present the results of the thermal conductivity $\kappa(\omega)$
as a function of frequency at doping $x=0.06$ (solid line),
$x=0.10$ (dashed line), and $x=0.12$ (dash-dotted line) for
parameter $t/J=2.5$ with temperature $T=0.05J$. Although
$\kappa(\omega)$ is not observable from experiments, its features
will have observable implications on the observable $\kappa(T)$.
From Fig. 1, we find the thermal conductivity spectrum consists of
two bands separated at $\omega\sim 0.5t$, the higher-energy band,
corresponding to the midinfrared band in the optical conductivity,
shows a weak peak at $\omega\sim 1t=2.5J$, while the position of
the lower-energy peak in the present underdoped cuprates is doping
dependent, and is located at a finite energy $\omega\sim xJ$.
Moreover, we also find from the above calculations that although
both dressed holons and spinons are responsible for the thermal
conductivity $\kappa(\omega)$, the contribution from the dressed
spinons is much larger than these from the dressed holons, i.e.,
$\kappa_{s}(\omega)\gg\kappa_{h} (\omega)$ in the underdoped
regime, and therefore the thermal conductivity of the underdoped
cuprates is mainly determined by its dressed spinon part
$\kappa_{s}(\omega)$. For a better understanding of the heat
transport of the underdoped cuprates, we have studied the
frequency dependence of the thermal conductivity spectrum at
different temperatures, and the results at $x=0.10$ for $t/J=2.5$
in $T=0.05J$ (solid line), $T=0.25J$ (dashed line), and $T=0.5J$
(dash-dotted line) are plotted in Fig. 2. These results show that
the high-energy band in $\kappa(\omega)$ is severely suppressed
with increasing temperatures, and vanishes at higher temperature
($T>0.4J$).
\begin{figure}[prb]
\epsfxsize=3.0in\centerline{\epsffile{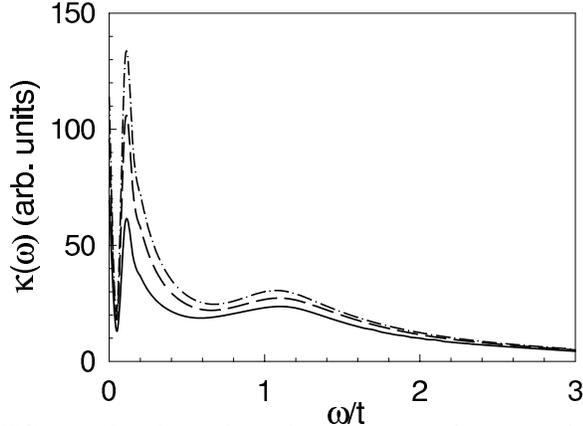}} \caption{The
thermal conductivity as a function of frequency at $x=0.06$ (solid
line), $x=0.10$ (dashed line), and $x=0.12$ (dotted line) with
$t/J=2.5$ in $T=0.05J$.}
\end{figure}
\begin{figure}[prb]
\epsfxsize=3.0in\centerline{\epsffile{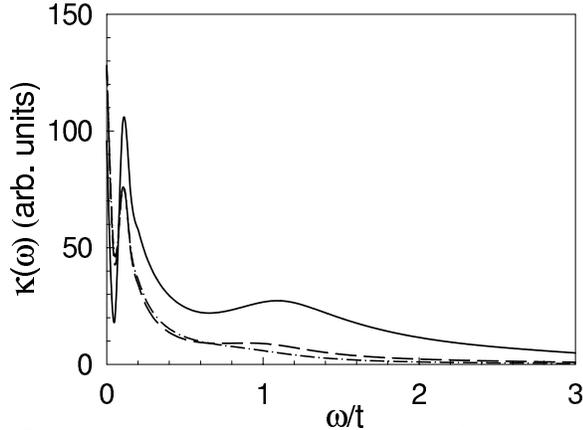}} \caption{The
thermal conductivity as a function of frequency at $x=0.10$ in
$T=0.05J$ (solid line), $T=0.25J$ (dashed line), and $T=0.5J$
(dotted line) with $t/J=2.5$.}
\end{figure}
\begin{figure}[prb]
\epsfxsize=3.0in\centerline{\epsffile{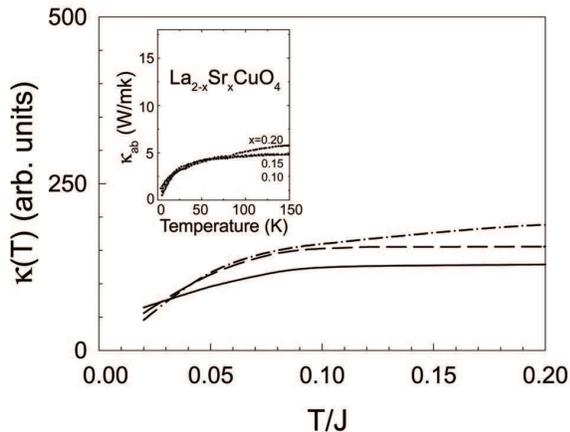}} \caption{The
thermal conductivity as a function of temperature at $x=0.10$
(solid line), $x=0.12$ (dashed line), and $x=0.15$ (dotted line)
with $t/J=2.5$. Inset: the experimental result of
La$_{2-x}$Sr$_{x}$CuO$_{4}$ taken from Ref. [9].}
\end{figure}

Now we turn to discuss the temperature dependence of the thermal
conductivity $\kappa(T)$, which can be obtained from Eq. (7) as
$\kappa(T)=\lim_{\omega\rightarrow 0}\kappa(\omega,T)$. The
results of $\kappa(T)$ at $x=0.10$ (solid line), $x=0.12$ (dashed
line), and $x=0.15$ (dash-dotted line) for $t/J=2.5$ are shown in
Fig. 3 in comparison with the experimental results \cite{nakamura}
taken on La$_{2-x}$Sr$_{x}$CuO$_{4}$ (inset). Our results show
that the thermal conductivity $\kappa(T)$ increases monotonously
with increasing temperatures for $T\leq xJ$, and is very weak
temperature dependent for $T> xJ$, in good agreement with the
experimental data \cite{nakamura} in the {\it normal state}, where
there is a shoulder around the temperature $T\sim xJ$ for both
experimental and theoretical results, and is consistent with the
position of the lower-energy peak in $\kappa(\omega)$. On the
other hand, we have noted that the smooth evolution of $\kappa(T)$
from the {\it superconducting} state ($T\leq 0.02J\approx 20$K) to
the {\it normal state} ($T> 0.02J$) has been observed from the
experiment \cite{nakamura} for La$_{2-x}$Sr$_{x}$CuO$_{4}$. Based
on the superconducting mechanism driven by the kinetic energy
\cite{feng3}, the thermal conductivity in the superconducting
state ($T\leq 0.02J \approx 20$K) is under investigation now.

In the above discussions, the central concern of the thermal
conductivity in the underdoped cuprates is the charge-spin
separation, then the heat transport is mainly determined by the
contribution from dressed spinons $\kappa_{s}(\omega,T)$. Since
$\kappa_{s}(\omega,T)$ in Eq. (7c) is obtained in terms of the
dressed spinon Green's function $D(k,\omega)$, while this dressed
spinon Green's function is evaluated by considering the
second-order correction due to the dressed holon pair bubble
\cite{feng1,feng5}, therefore the observed unusual frequency and
temperature dependence of the thermal conductivity spectrum of the
underdoped cuprates is closely related to the incommensurate spin
dynamics \cite{cheng}. This is because that within the
fermion-spin theory, the dynamical spin structure factor has been
obtained \cite{feng1,feng5} in terms of the full dressed spinon
Green's function as,
\begin{eqnarray}
S({\bf k},\omega)&=&-2[1+n_{B}(\omega)]{\rm Im}D(k,\omega)\nonumber \\
&=&[1+n_{B}(\omega)]A_{s}(k,\omega) \nonumber \\
&=&-2[1+n_{B}(\omega)]B_{k}{\rm Im}\Sigma_{s}({\bf k},
\omega)\over [\omega^{2}-\omega^{2}_{k}-{\rm Re} \Sigma_{s} ({\bf
k},\omega)]^{2}+[{\rm Im} \Sigma_{s}({\bf k},\omega)]^{2},
\end{eqnarray}
where ${\rm Im}\Sigma_{s}({\bf k},\omega)$ and ${\rm Re}
\Sigma_{s}({\bf k},\omega)$ are corresponding imaginary part and
real part of the dressed spinon self-energy function
$\Sigma_{s}({\bf k},\omega)$ in Eq. (8b). As we have shown in
detail in Refs. \cite{feng1,feng5}, the dynamical spin structure
factor (9) has a well-defined resonance character. $S({\bf
k},\omega)$ exhibits a peak when the incoming neutron energy
$\omega$ is equal to the renormalized spin excitation
$E^{2}_{k}=\omega^2_{k}+B_{k}{\rm Re} \Sigma_{s}(k,E_{k})$ for
certain critical wave vectors ${\bf k}_{\delta}$ (positions of the
incommensurate peaks). The height of these peaks is determined by
the imaginary part of the dressed spinon self-energy $1/{\rm Im}
\Sigma_{s}({\bf k}_{\delta}, \omega)$, i.e., the height of the
incommensurate peaks is determined by damping, it then is fully
understandable that they are suppressed as the energy and
temperature are increased. Since this incoming neutron resonance
energy $\omega=E_{k}$ is finite, this leads to that the
lower-energy peak in $\kappa(\omega)$ is located at a finite
energy. Near the half-filling, the spin excitations are centered
around the AF wave vector $[\pi,\pi]$, so the commensurate AF peak
appears there. Upon doping, the dressed holons disturb the AF
background. Within the fermion-spin framework, as a result of
self-consistent motion of the dressed holons and spinons, the
incommensurate antiferromagnetism is developed away from the
half-filling, where the incommensurate peaks are located at
$[(1\pm\delta)\pi,\pi]$ and $[\pi,(1\pm\delta)\pi]$ with the
incommensurability parameter $\delta(x)$ defined as the deviation
of the peak position from the AF wave vector $[\pi,\pi]$. This
incommensurability parameter $\delta(x)$ increases progressively
with the doping concentration at lower dopings \cite{feng1,feng5},
which leads to that the position of the lower-energy peak in
$\kappa(\omega)$ is doping dependent. Using a typical value of
$\delta$ in the underdoped regime obtained in the previous
calculations \cite{feng1,feng5}, the distance of the
incommensurate peaks is estimated as 20\AA. On the other hand,
using a typical velocity of sound in La$_{2-x}$Sr$_{x}$CuO$_{4}$
of about 5km/s and a phonon energy of 10 mev, one \cite{baberski}
obtains a wavelength for the phonons of 15\AA. This value is in
qualitative agreement with the distance of the incommensurate
peaks, and therefore the time scale of the dynamic incommensurate
correlation is comparable to that of the lattice vibrations
\cite{baberski}. In this case, the dynamic lattice modulations are
induced, then the dynamic spinon modulations dominate the heat
transport of the underdoped cuprates, in other words, the
incommensurate spin fluctuation has been reflected in the thermal
conductivity. On the other hand, these incommensurate peaks are
very sharp at low temperatures and energies \cite{feng1,feng5},
however, they broaden and weaken in amplitude as the energy
increase for low energies $\omega \leq xJ$, and almost vanishes
for high energies $\omega > xJ$, which leads to the shoulder
appears in $\kappa(T)$ in the temperature $T\sim xJ$. Finally, we
emphasize that the present theory can describe the heat transport
of the underdoped single layer cuprates, where only incommensurate
spin fluctuation is observed \cite{cheng}. However, both
incommensurate spin fluctuation and commensurate $[\pi,\pi]$
resonance have been observed in the underdoped bilayer cuprates in
the normal state \cite{dai} due to the bilayer splitting in the
band structure \cite{feng9}. This resonance may lead some
additional features in the thermal conductivity of the underdoped
bilayer cuprates \cite{takenaka}, and these and other related
issues are under investigation now.

In summary, we have studied the heat transport of the underdoped
single layer cuprates within the $t$-$J$ model. Our results show
that the frequency dependence of the thermal conductivity spectrum
$\kappa(\omega)$ at low temperatures consists of two bands. The
high energy band shows a weak peak, while the low energy peak is
located at a finite energy. This high energy broad band is
severely suppressed with increasing temperatures, and vanishes at
higher temperature. Moreover, the temperature dependence of the
thermal conductivity $\kappa(T)$ increases monotonously with
increasing temperatures. Our results also show that although both
dressed holons and spinons are responsible for the thermal
conductivity, the contribution from the dressed spinons dominates
the heat transport of the underdoped cuprates. On the other hand,
we emphasize that although the simplest $t$-$J$ model can not be
regarded as a complete model for the quantitative comparison with
the doped cuprates, our present results of the thermal
conductivity are in qualitative agreement with the major
experimental observations of the underdoped single layer cuprates
\cite{nakamura}.

\acknowledgments The authors would like to thank Dr. Jihong Qin
for the helpful discussions. This work was supported by the
National Natural Science Foundation of China under Grant Nos.
10125415 and 90103024, and the Grant from Beijing Normal
University.

\end{document}